\documentclass[aps,prl,twocolumn,showpacs,amsmath,amssymb,superscriptaddress]{revtex4}
\usepackage{amsfonts}
\usepackage{amsmath}
\usepackage{dcolumn}
\usepackage[dvips]{graphicx,color}
\begin{document}
\title{Dynamical Mean-Field Study of the Ferromagnetic Transition Temperature
of a \\
Two-Band Model for Colossal Magnetoresistance Materials}

\author{F. Popescu}
%\footnote{florentin\_p@hotmail.com}
\affiliation{Department of Physics, Florida State University,
Tallahassee, FL 32306}

\author{C. \c{S}en}
%\footnote{sen@magnet.fsu.edu}
\affiliation{Department of Physics, Florida State University,
Tallahassee, FL 32306}

\author{E. Dagotto}
%\footnote{edagotto@utk.edu}
\affiliation{Department of Physics and Astronomy, University of
Tennessee, Knoxville, TN 37996} \affiliation{Condensed Matter
Sciences Division, Oak Ridge National Laboratory, Oak Ridge, TN
32831}

\begin{abstract}
The ferromagnetic (FM) transition temperature ($T_{\rm C}$) of a
two-band Double-Exchange (DE) model for colossal magnetoresistance
(CMR) materials is studied using dynamical
mean-field theory (DMFT), in wide ranges of coupling constants,
hopping parameters, and carrier densities. The results are shown to
be in excellent agreement with Monte Carlo simulations. When the
bands overlap, the value of $T_{\rm C}$ is found to be much larger
than in the one-band case, for all values of the chemical potential
within the energy overlap interval. A nonzero interband hopping
produces an additional substantial increase of $T_{\rm C}$, showing
the importance of these nondiagonal terms, and the concomitant use
of multiband models, to boost up the critical temperatures in
DE-based theories.
\end{abstract}

\pacs{75.47.Lx, 71.27.+a, 71.30.+h}

\maketitle

The study of the CMR rare-earth perovskites has recently attracted
considerable attention due to the rich magnetic and structural
transitions they display \cite{URU95}. The CMR effect, namely an
extremely large drop in resistivity caused by a magnetic field,
occurs at the transition between a low-temperature FM-metallic
ground state and a high-temperature paramagnetic-insulating phase,
i.e., near $T_{\rm C}$. A basic aspect of CMR physics is that the
mobile carriers (Mn $e_{\rm g}$-symmetric electrons) are strongly
coupled ferromagnetically to localized spins (Mn $t_{\rm
2g}$-symmetric electrons). The core spin alignment influences on the
electron motion leading to ferromagnetism. The basic model used to
describe CMR materials is the DE model \cite{ZEN51}, formulated with
one localized spin per site coupled to mobile carriers via a Hund's
coupling $J$ much larger than the hopping amplitudes. While DE ideas
explain qualitatively the existence of ferromagnetism, much of the
CMR physics is still under debate \cite{DAG05,MAN05}. This is in
part caused by the absence of fully reliable many-body techniques to
study the complicated DE models needed to describe these compounds.
Only MC and static mean-field approximations have been applied to
the realistic multiband problem \cite{DAG05}, the DMFT treatment
being notoriously absent. Despite a considerable effort carried out
within DMFT in the context of one-band models \cite{FUR94}, the
realistic case of two active bands has received much less attention
since its analysis is far more difficult.

In this paper, the DMFT method is applied for the first time to the
analysis of $T_{\rm C}$ of a two-band model for CMR materials. Its
results are contrasted against MC simulations and found to be in
good agreement. Besides the relevance of this combined DMFT+MC
contribution, which is highly nontrivial technically as shown below,
we also unveil here the key relevance of the \textit{interband}
hopping amplitudes to boost the value of the critical FM
temperature. This effect was also not discussed in previous
literature, and may lead to creative procedures to further increase
$T_{\rm C}$ in real materials. The DMFT approach is general, with
two $s$=$1/2$ active bands, arbitrary couplings, hoppings, and
carrier densities $p$.

The two-band model is based on the DE Hamiltonian:
\begin{equation}\label{ham}
{\mathcal{H}}\!=-\!\!\!\sum_{ll^{'},\langle ji\rangle,\alpha}\!\!
(t_{ll^{'}}c^{\dag}_{l^{'},j,\alpha}c_{l,i,\alpha}\!+\textrm{H.c.})
-2\sum_{l,i} J_l \mathbf{S}_{i}\cdot\mathbf{s}_{l,i},
\end{equation}
where $l,l^{'}$ ($=1,2$) are the $e_{\rm g}$-band indexes, $i,j$
label the sites, $c_{l,i,\alpha}$ destroys an electron at site $i$
in the band $l$,
$\mathbf{s}_{l,i,\beta\alpha}=c^{\dag}_{l,i,\beta}(\mathbf{\sigma}_{\beta\alpha}/2)c_{l,i,\alpha}$
is the spin- operator of the mobile carrier at band $l$
($\hat{\mathbf{\sigma}}$ = Pauli vector), $\alpha$ and $\beta$ are
spin indexes, $J_{l}$ is the coupling between the core spin and the
conduction electrons of band $l$, and $\mathbf{S}_{i}$ is the
localized spin of the magnetic ion at site $i$. For $l=l^{'}$ we
refer to $t_{ll}$ as the band hopping ($\equiv t_{l}$). For $l\neq
l^{'}$, the $t_{ll^{'}}$ is referred to as the interband hopping
($t_{ll^{'}}=t_{l^{'}l}$). The two active bands couple through the
simultaneous scattering of carriers with the same core spin, as well
as through the exchange of carriers via the nondiagonal hopping.
While the first coupling clearly causes an \textit{increase} in
$T_{\rm C}$ when the bands overlap within the same energy interval,
namely when $J_{l}$ are closed, the electron exchange among bands is
{\it also} shown to induce higher $T_{\rm C}$ even when the
$J_{l}$'s are very different.

Note that the addition of other important terms, such as the
antiferromagnetic exchange $J_{\rm AF}$ between core spins and/or
the cooperative Jahn-Teller phonons, will need  a
sophisticated ``cluster'' DMFT where at least some short-distance
effects are considered. For this reason, this first study using the
DMFT method on a two-band model focuses on the simplest case where
only the FM phase is relevant. The study of the competition with
other phases is left for future investigations. The model is studied
below using DMFT and MC techniques.

{\it DMFT results:} Within DMFT, the self-energy is local;
$\Sigma(\mathbf{p},i\omega_{n})$$\rightarrow$$\Sigma(i\omega_{n})$
($\omega_{n}$=$(2n\!+\!1)\pi T$ are the Matsubara frequencies).
Since $\Sigma$ is momentum independent, the information about the
hopping of carriers on and off lattice sites is carried by the bare
Green's function ${\mathcal{G}}_{0}(i\omega_{n})$.  With the
effective action ${\mathcal{S}}_{eff}(\mathbf{m})$ defined by
${\mathcal{G}}_{0}$ quadratic in the Grassman variables, the full
Green's function ${\mathcal{G}}$ can be solved by integration:
$\langle{\mathcal{G}}(i\omega_{n})\rangle=\langle[{\mathcal{G}^{-1}_{0}}
(i\omega_{n})+J\mathbf{S}{\mathbf{m}}\hat{\sigma}]^{-1}\rangle$. The
average $\langle X(\mathbf{m})\rangle=\int
d\Omega_{m}X(\mathbf{m})P(\mathbf{m})$ is over the orientations
$\mathbf{m}$ of the local moment with the probability
$P(\mathbf{m})$. In the FM phase near $T_c$,
$P(\mathbf{m})\propto\exp{(-3\beta M\mathbf{m})}$, where
$\beta=1/T$, and $M=\langle m_{z}\rangle_{\mathbf{m}}$ is the
local-order parameter, nonzero only below $T_{\rm C}$. If two bands
are active, then one rewrites the average above for each band;
$\langle{\mathcal{G}}_{l}(i\omega_{n})\rangle=\langle[{\mathcal{G}^{-1}_{0,l}}
(i\omega_{n})+J_{l}\mathbf{S}_{l}{\mathbf{m}}\hat{\sigma}]^{-1}\rangle$,
and solve the equations for the coupled Green's functions;
$\langle{\mathcal{G}}_{0,l}^{-1}(i\omega_{n})\rangle=z_{n}-t^{2}_{l}
\langle{\mathcal{G}}_{l}(i\omega_{n})\rangle-t^{2}_{ll^{'}}
\langle{\mathcal{G}}_{l^{'}}(i\omega_{n})\rangle$, ($l\neq l^{'}$)
on a Bethe lattice with a semicircular noninteracting
Density-Of-States
$\mathrm{DOS}_{l}(\omega)=(4t_{l}^{2}-\omega^{2})^{1/2}/2\pi
t^{2}_{l}$, where $z_{n}=i\omega_{n}+\mu$ ($\mu$ = chemical
potential). $t_{ll^{'}}$ carries in ${\mathcal{G}}_{0,l}$ the
information about the second band $l^{'}$ through
${\mathcal{G}}_{l^{'}}$. To find $T_{\rm C}$, we parametrize
${\mathcal{G}}^{-1}_{0,l}$:
${\mathcal{G}}^{-1}_{0,l}(i\omega_{n})=[z_{n}+R_{l}(i\omega_{n})]
\hat{\mathbf{1}}+Q_{l}(i\omega_{n})\hat{\mathbf{\sigma}}_{z}$, and
linearize ${\mathcal{G}}^{-1}_{0,l}$ with $M$. Up to first order:
\begin{equation}
R_{l}=-t^{2}_{l}\frac{B_{l}}{B_{l}^{2}-J^{2}_{l}}
-t^{2}_{ll^{'}}\frac{B_{l^{'}}}{B_{l^{'}}^{2}-J^{2}_{l^{'}}},\label{EQ1}
\end{equation}
where $B_{l}(i\omega_{n})=z_{n}+R_{l}(i\omega_{n})$. A similar
equation for $R_{l^{'}}$ can be obtained by interchanging the band
indexes $l\rightarrow l^{'}$. Despite the extreme complexity of
Eqs.~(\ref{EQ1}), we managed to perform analytical calculations to
decouple the equations set and to obtain separate equations for
$R_{l}$ and $R_{l^{'}}$. The  polynomial coefficients of the $9$-th
order equations for $R_{l}$ and $R_{l^{'}}$ which were solved
numerically, are combinations of all parameters of the model, i.e.,
couplings and hoppings (not reproduced here because of their size).
For $t_{ll^{'}}=0$, the $9$-th order equations reduce to the $3$-th
order ones:
$R^{3}_{l}+2z_{n}R^{2}_{l}+(z_{n}^{2}+t^{2}_{l}-J^{2}_{l})R_{l}+z_{n}t^{2}_{l}=0$.
At $\mu=0$ and with the substitution $i\omega_{n}\rightarrow\omega$,
Eq.\ (\ref{EQ1}) gives the interacting electronic $\mathrm{DOS}_{l}$
at $T=0$ \cite{DOS}. Due to having a nonzero $t_{ll^{'}}$, besides
its own parameters $t_{l}$ and $J_{l}$, $\mathrm{DOS}_l$ depends
also on the parameters characterizing the other band $l^{'}$, i.e.,
$t_{l^{'}}$ and $J_{l^{'}}$. This interplay of parameters describing
the coupling of bands leads to an increase in $T_{\rm C}$ even when
the $J_{l}$'s are very much different, i.e. when the bands occupy
different energy intervals. The obtained equation for
$Q_{l}(i\omega_{n})$:
\begin{eqnarray}
Q_{l}&=&t^{2}_{l}\frac{J_{l}M+Q_{l}}{{B_{l}^{2}-J^{2}_{l}}}+t^{2}_{ll^{'}}
\frac{J_{l^{'}}M+Q_{l^{'}}}{{B_{l^{'}}^{2}-J^{2}_{l^{'}}}}\nonumber\\
&+&t^{2}_{l}\frac{2J^{2}_{l}Q_{l}}{3({B_{l}^{2}-J^{2}_{l}})^{2}}
+t^{2}_{ll^{'}}\frac{2J^{2}_{l^{'}}Q_{l^{'}}}{3({B_{l^{'}}^{2}-J^{2}_{l}})^{2}},\label{EQ2}
\end{eqnarray}
leads to an implicit expression for $T_{\rm C}$ in the form:
\vspace{-0.35in}
\begin{widetext}
\begin{equation}
-\frac{4}{3}\sum^{\infty}_{n=0}\frac{\sum_{l}t^{2}_{l}J^{2}_{l}(B^{2}_{l}-J^2_{l})^{2}
+2t^{2}_{ll^{'}}\prod_{l}J_{l}(B^{2}_{l}-J^2_{l})
-(t^{2}_{l}t^{2}_{l^{'}}-t^{4}_{ll^{'}})\sum_{l}J^{2}_{l}(B^{2}_{l}-J^{2}_{l}/3)}
{\prod_{l}(B^{2}_{l}-J^{2}_{l})^{2}-\sum_{l}t^{2}_{l}(B^{2}_{l}
-J^{2}_{l}/3)(B^{2}_{l^{'}}-J^2_{l^{'}})^{2}
+(t^{2}_{l}t^{2}_{l^{'}}-t^{4}_{ll^{'}})\prod_{l}
(B^{2}_{l}-J^{2}_{l}/3)}=1,\label{TC}
\end{equation}
\end{widetext}
where above, if $l=1$ (2), then $l^{'}=2$ (1). At $t_{ll^{'}}=0$,
Eq.(\ref{TC}) reduces to:
\begin{equation}
\sum^{2}_{l=1}\sum_{n=0}^{\infty}\frac{-2t^{2}_{l}J^{2}_{l}}{3(B^{2}_{l}-J^{2}_{l})^{2}-3t^{2}_{l}
(B^{2}_{l}-J^{2}_{l})-2t^{2}_{l}J^{2}_{l}}=1,\label{CUBTC}
\end{equation}
where $B_{l}$ satisfies: $B^{3}_{l}-z_{n}B^{2}_{l}+(t^{2}_{l}
-J^{2}_{l})B_{l}+z_{n}J^{2}_{l}=0$. We tested Eqs.~(\ref{TC}),
(\ref{CUBTC}) in several cases: (1) at $t_{2}$=$t_{12}=0$ and
$J_{2}$=$0$ we recovered the one-band model results reported in
Ref.~\cite{AUS01}; (2) at $t_{2}$=$t_{12}$=$0$, $J_{2}$=$0$ and $J_1
\rightarrow \infty$, the results of Ref.~\cite{FIS03} are
reproduced. Details on the calculations above will be published
elsewhere. From Eq.~(\ref{TC}), the $T_{\rm C}$ contained in the
Matsubara frequencies was extracted numerically. Eq.~(\ref{CUBTC})
predicts an increase in $T_{\rm C}$ only for the values of $\mu$
within the energy overlap interval.

\begin{figure}[t]
{\scalebox{0.32}{\includegraphics[clip,angle=0]{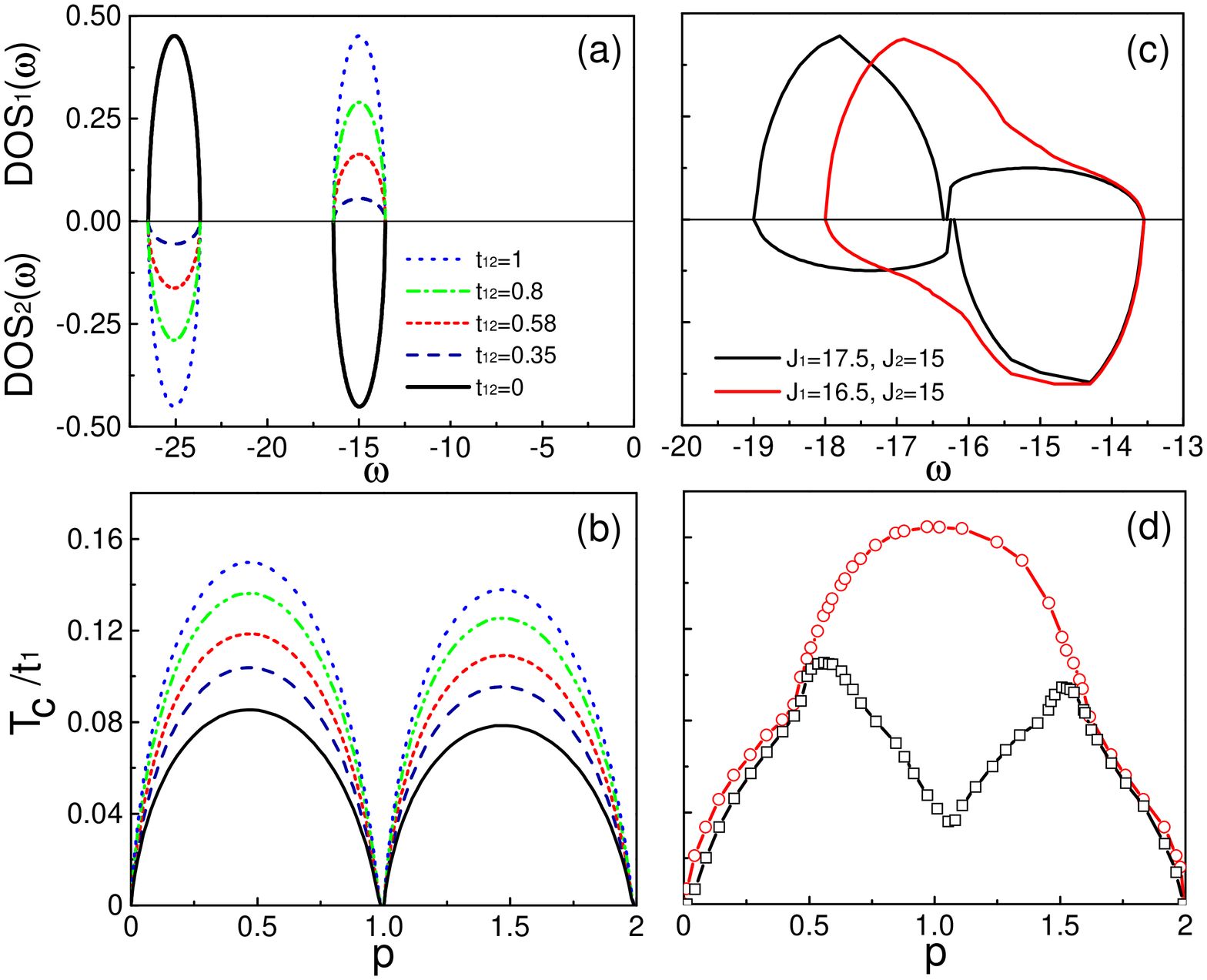}}}
\caption{ (a) DMFT zero-temperature interacting $\mathrm{DOS}_{l}$ for
$J_{1}$=$25$ and $J_{2}$=$15$ at different values of $t_{12}$; (b)
$T_{\rm C}$ vs. $p$\, for the cases shown in (a). Since the Hund's
couplings are very different, the bands and critical temperatures
retain the semicircular form; (c) Zero-temperature interacting
$\mathrm{DOS}_{l}$ at $t_{12}$=$0.5$, for the couplings indicated;
(d) $T_{\rm C}$ vs. $p$\, for the cases shown in (c). Even if the
overlap of bands is narrow, the exchange of electrons is strong
enough to induce a robust $T_{\rm C}$ at $p=1$ (black curve). In all
frames $t_{1}$=$t_{2}$=$1$.}\label{Fig1}
\end{figure}

In Fig.\ \ref{Fig1}(a), the $\mathrm{DOS}_{l}(\omega)$ at
$J_{1}/t_{1}$=$25$ and $J_{2}/t_{2}$=$15$ is shown for several
$t_{ll^{'}}$. Our $J_{l}/t_{l}$ are much larger than the value
$(J_{l}/t_{l})_{\rm min}\approx 1.4$ that corresponds to the
electron and hole bands formation, and than the value $J/t$=$8$
considered in the one-band model large enough to capture properly
the FM features of CMRs \cite{DAG03}. Hence, the electron bands are
centered at $\omega^{-}_{1}$=$-25$ and $\omega^{-}_{2}$=$-15$,
respectively (the hole bands, centered at $\omega^{+}_{1}$=$25$ and
$\omega^{+}_{2}$=$15$ due to the electron-hole symmetry, are not
shown here for simplicity). For clarity, we plotted
$\mathrm{DOS}_{2}(\omega)$ with a reversed sign. Since
$|J_{1}/t_{1}-J_{2}/t_{2}|\gg 1$, the bands occupy different energy
intervals. As a consequence, each band gives its own contribution to
$T_{\rm C}$ (Fig.\ \ref{Fig1}(b)). At $t_{ll^{'}}$=$0$ the bands are
fully decoupled, and the values of $T_{\rm C}$ match the results of
the one-band model for all $p$'s. However, if $t_{ll^{'}}\neq 0$,
then the carriers are allowed to hop between the bands and, thus,
they can belong simultaneously to both bands. This leads to an
increase in the effective number of interacting electrons of each
active band. As shown in Fig.~\ref{Fig1}(a), due to the transfer of
electrons among bands, in $\mathrm{DOS}_{l}(\omega)$ a new region
occupied by the interacting electrons builds up within the interval
of energies occupied by the $\mathrm{DOS}_{l^{'}}(\omega)$. At
$t_{ll^{'}}$=$t_{l}$, the effective number of interacting electrons
within each energy interval becomes twice as large. In consequence,
the $T_{\rm C}$ for all $p$ corresponding to $\mu$ within each
energy interval, is almost twice larger than in the one-band case
(Fig.\ \ref{Fig1}(b)). By contrary, when the $J_{l}$'s are similar
the coupling of bands is strong enough to lead to a deviation from
the semicircular form, especially when $t_{ll'}\sim
|J_{l}-J_{l^{'}}|$ (Fig.\ \ref{Fig1}(c)). If the bands partially
overlap, a hump develops in $T_{\rm C}$ at all values of $p$
corresponding to $\mu$ within the energy overlap interval (Fig.\
\ref{Fig1}(d)).

\begin{figure}[t]
{\scalebox{0.33}{\includegraphics[clip,angle=0]{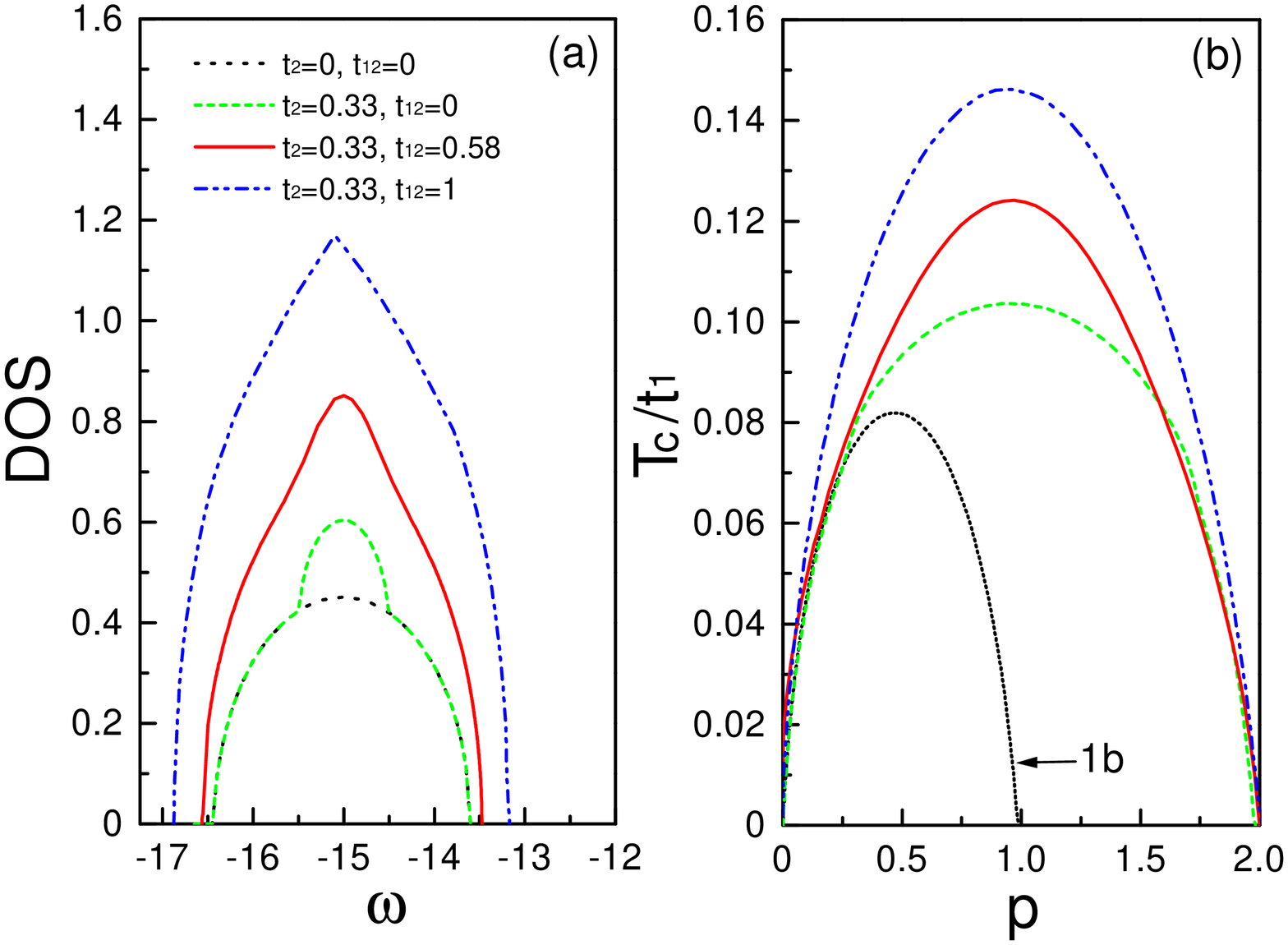}}}
\caption{(a) DMFT zero-temperature total interacting DOS for different
values of $t_{12}$. (b) $T_{\rm C}$ vs. $p$ for the parameters
indicated in (a). The red curve corresponds to the realistic set of
hoppings for $\mathrm{LaMnO_{3}}$. In all frames $t_{1}$=$1$,
$J_{1}$=$J_{2}$=$15$. The black-dot curve corresponds to the
one-band case.}\label{Fig2}
\end{figure}

In Fig\ \ref{Fig2}(a), the total interacting $DOS(\omega)$
\cite{DOS} at $t_{1}$=$1$, $t_{2}$=$1/3\sim0.33$ is shown, for
different $t_{ll^{'}}$ in the case when the bands fully overlap
($J_{1}/J_{2}$=$1$). The electron exchange effect does not only
increase the effective number of interacting electrons in each band,
but also extends the energy region occupied by the bands. However,
the total number of interacting electrons does not change, the bands
being fully filled for $p$=$2$. At $t_{12}$=$0$ (Eq.\ (\ref{CUBTC}))
the increase in $T_{\rm C}$ is due to the overlap of bands (green
dashed curve in Fig.2(b)). However, when $t_{ll^{'}}\neq 0$ (see
Eq.~(\ref{TC})) the exchange effect further boosts $T_{\rm C}$
(Fig.\ \ref{Fig1}(b)).

{\it Monte Carlo results:} The Hamiltonian (\ref{ham}) was also
studied numerically using the MC methods widely applied to Mn-oxides
\cite{DAG01} in the limit $J\rightarrow\infty$ \cite{DAG03}. Hence,
the $e_{\rm g}$-spins are perfectly aligned with the $t_{\rm 2g}$
spin. The technique involves finding the eigenvalues of the
Hamiltonian matrix at each MC step corresponding to a newly updated
set of localized spins. Although this substantially limits the size
of the clusters being simulated, the results for small lattices are
numerically exact and they allow for a direct comparison with DMFT.
We have simulated lattices of sizes 8$\times$8 in 2D and
4$\times$4$\times$4 in 3D. The core spins are treated classically,
while the treatment of the fermionic sector is exact. In all
simulations $2\times10^{4}$ MC steps were used, the first $10^4$
been discarded in order to account properly for the thermalization
of the random starting configuration. In finite dimensions, the
hopping carries a direction index ``$\alpha$", that is
$t_{l,l^{\prime}}^{\alpha}$. In 2D (3D),
$t_{l,l}^{x}$=$-\sqrt{3}t_{l,l^{\prime}}^{x}$=$-\sqrt{3}t_{l^{\prime},l}^{x}$=$3t_{l^{\prime},l^{\prime}}^{x}=1$,
$t_{l,l}^{y}$=$\sqrt{3}t_{l,l^{\prime}}^{y}$=$\sqrt{3}t_{l^{\prime},l}^{y}
=3t_{l^{\prime},l^{\prime}}^{y}=1$,
($t^{z}_{l,l}$=$t^{z}_{l,l^{\prime}}$=$t^{z}_{l^{\prime},l}$=$0$,
$t_{l^{\prime},l^{\prime}}^{z}$=$4/3$), in the $x$-, $y$- (and $z$-)
directions, respectively \cite{SLA54}. The indexes $l$ and $l^{'}$
stand for the two active, $x^{2}-y^{2}$ and $3z^{2}-r^{2}$,
orbitals. $t_{1}$=$1$ sets the energy unit. To find $T_{\rm C}$ we
investigate the long-range spin-spin correlations:
\begin{equation}
S(x)=\!\frac{1}{N}\sum_{i}\langle
\vec{S}_i\cdot\vec{S}_{i+x}\rangle=\!\frac{1}{N}
\sum_{i}\frac{\mbox{Tr}(\vec{S}_{i}\vec{S}_{i+x}e^{-\beta
{\mathcal{H}}})} {\mbox{Tr}(e^{-\beta {\mathcal{H}}})},
\end{equation}
where $N$ is the total number of sites. $T_{\rm C}$ is the
temperature for which $S\rightarrow 0$ upon heating, at the maximum
distance $x_{max}$ in the clusters considered.

\begin{figure}[t]
{\scalebox{0.5}{\includegraphics[clip,angle=0]{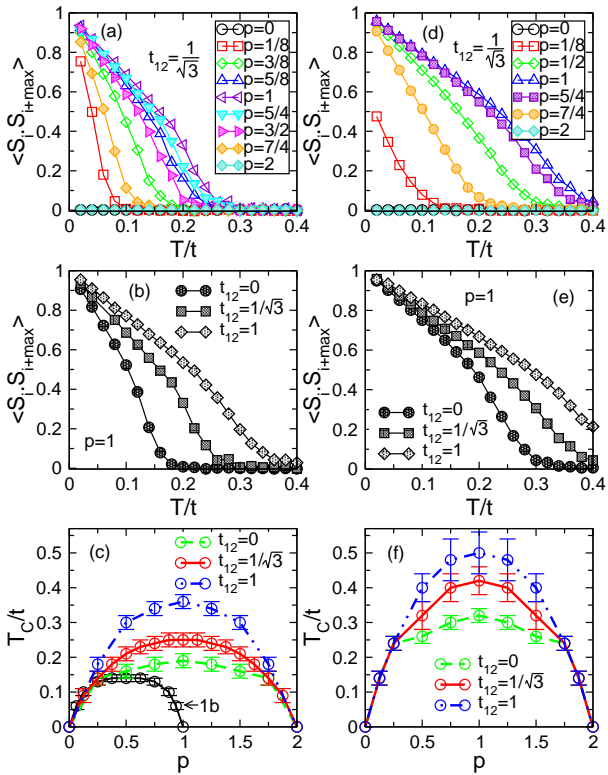}}}
\caption{(a) MC spin-spin correlations at the maximum distance
$4\sqrt{2}$ vs. temperature for different values of $p$, at
$t_{1}$=$1(\equiv t)$, $t_{2}$=$1/3$, and the $t_{12}$ indicated,
using a 2D lattice of size 8$\times$8. Errors are less than the
symbol size. (b) Same as in (a), but changing $t_{12}$; (c) $T_{\rm
C}$ vs. $p$ at different $t_{12}$ for the same lattice used in (a).
The black line corresponds to the one-band case. The red line shows
the $T_{\rm C}$ obtained from (a). (d-f) Same as in (a-c), but using
a 3D lattice of size 4$\times$4$\times$4. In all frames
$J_{1}=J_{2}\rightarrow\infty$.} \label{Fig3}
\end{figure}

In Fig.\ \ref{Fig3}, the results in 2D and 3D are displayed
side-by-side for comparison.
In panel (a), the spin-spin correlations at $x$=$x_{max}$ is
shown for an 8$\times$8 lattice, corresponding to different electron
densities $p$=$N_{e}/2N$, where $N_e$ is the total number of
electrons. The same in (d), but on a $4^{3}$ lattice. Moreover, we
investigated $T_{\rm C}$ vs. $p$ at $t_1$=$1$, $t_2$=$1/3$, and
different $t_{12}$. The results are in Fig.\ \ref{Fig3}(b) for 2D
and in Fig.\ \ref{Fig3}(e) for 3D. The $T_C$ is maximum at $p$=$1$
and vanishes in the limits $p$$\rightarrow$$0$ and
$p$$\rightarrow$$2$, with an overall semicircular form in the phase
diagram, as seen in the frames (c) and (f). The one-band phase
diagram (denoted 1b) is also shown. Hence, as seen in (c) and (f),
$t_{12}$ has a substantial effect in rising the $T_{\rm C}$, which
is in qualitative, and even in quantitative \cite{CUANT}, agreement
with DMFT. As $t_{12}$ increases from $0$ to $1$, the increase in
$T_C$ can be as high as $100$\% in 2D and $60$\% in 3D.
\begin{figure}[t]
{\scalebox{0.3}{\includegraphics[clip,angle=0]{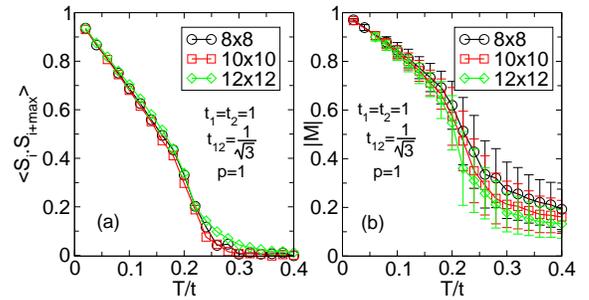}}}
\caption{MC finite-size effects studied in 2D dimensions. (a)
Spin-spin correlations at the maximum allowed distance vs.
temperature for the lattice sizes indicated. (b) Magnetization $|M|$
vs. temperature corresponding to the same parameters in (a). The
non-zero value of $|M|$ at high temperatures is the asymptotic value
$1/\sqrt{N}$ for a system of size $N$. 3D results using
5$\times$5$\times$5 lattices are similar (not shown). While the
spin-spin correlations do not show appreciable size effects, the
magnetization results clearly do, their errors bars being maximal
within the critical region.}\label{Fig4}
\end{figure}

The finite-size effects are checked using different boundary
conditions as well as simulating clusters of up to 12$\times$12 in
2D and 5$\times$5$\times$5 in 3D. The results are in Fig.\
\ref{Fig4}, where the spin-spin correlations at $x=x_{max}$ (a) and
the magnetization $|M|$ vs. $T$ curves (b) are shown. The size
effects are small, showing that our MC method detects the
$T_{\rm C}$ accurately.

{\it Conclusion}: We carried out the first study of a multiband DE
model applied to CMR using a powerful combination of DMFT and MC
techniques. When two active bands are considered, the $T_{\rm C}$ is
maximized at $p$=$1$. DMFT shows that the interband hopping leads to
an increase in $T_{\rm C}$ at all $p$'s, even if the electron bands
do not occupy the same energy interval. This is due to the electron
exchange between the bands, which increases the effective number of
interacting electrons within each band. Both DMFT and MC indicate
that, if the bands fully overlap, besides the increase of $T_{\rm C}$
due to the energy overlap, a further boost occurs when the interband
hopping is turned on. The ideas developed here can be used to search
for multiorbital FM materials with even higher $T_{\rm C}$ than
currently known, once the interband hopping is tuned up. Our study
can be extended to Diluted Magnetic Semiconductors, with similar results
expected \cite{POP06}.

We acknowledge conversations with R.S. Fishman, J. Moreno, G.
Alvarez, A. Moreo, and T. Carsten. This research was supported by
grant NSF-DMR-0443144.

%\suppressfloats

\begin{references}
\bibitem{URU95} Urushibara {\it et al.}, Phys. Rev. B {\bf 51},
14103 (1995), Y. Tokura {\it et al.}, J. Phys. Soc. Jpn. {\bf 63},
3931 (1994).
\bibitem{ZEN51} C. Zenner, Phys. Rev. {\bf 82}, 403 (1951), P.W.
Anderson and H. Hasegawa, Phys. Rev. {\bf 100}, 657 (1955), P.G. De
Gennes, Phys. Rev. {\bf 118}, 141 (1960)
\bibitem{DAG05} For a discussion on open issues regarding CMR see: E Dagotto, New J. Phys.  {\bf
7}, 67 (2005).
\bibitem{MAN05} N. Mannella {\it et al.}, Nature, {\bf 438}, 474
(2005).
\bibitem{FUR94} N. Furukawa, J. Phys. Soc. Japan {\bf 63}, 3214
(1994), ibid. cond-mat/9812066.
\bibitem{DOS} $DOS_{l}(\omega)=-2Im{[R_{l}(\omega)]}/\pi$, $DOS(\omega)=\sum_{l}DOS_{l}(\omega)$.
\bibitem{AUS01} M. Auslender and E. Kogan,
Phys. Rev. B {\bf 65}, 012408 (2001); ibid. {\bf 67}, 132410,
(2003), M. Auslender and E. Kogan, Europhys. Lett. {\bf 59}, 277
(2002).
\bibitem{FIS03} R.S. Fishman and M. Jarrell,
J. Appl. Phys. {\bf 93}, 7148 (2003); ibid. Phys. Rev. B {\bf 67},
100403 (2003).
\bibitem{DAG03} For an overview of works on the manganites, see E.
Dagotto, {\it Nanoscale Phase Separation and Colossal
Magnetoresistance}, Springer, Berlin, 2003.
\bibitem{DAG01} E. Dagotto {\it et al.}, Phys. Rep. {\bf 344}, 1,
(2001), S. Yunoki {\it et al.}, Phys. Rev. Lett. {\bf 80}, 845,
(1998).
\bibitem{SLA54} J.C. Slater and G.F. Koster, Phys. Rev. {\bf 94},
1498 (1954).
\bibitem{CUANT} With a bandwidth $W=1$eV at $p=1$, DMFT ($W=4t$), at $J_{1}$=$J_{2}$=$15$, $t_{1}\cong 0.25$eV,
$t_{2}=t_{1}/3\cong 0.083$eV, $t_{12}=t_{1}/\sqrt{3}\cong 0.144$eV,
gives $T_{\rm C}\sim 370$K, while MC gives in 2D ($W=6t$) at $t_{1}$
$\sim0.17$eV the $T_{\rm C}\sim 425$K, and in 3D ($W=12t$) at
$t_{1}\sim 0.084$eV the $T_{\rm C}\sim 407$K.
\bibitem{POP06} F. Popescu {\it et al.}, Phys. Rev. B {\bf 73},
075206 (2006).
\end{references}
\end{document}